\documentclass[twocolumn,aps,floatfix,longbibliography]{revtex4-1}
\usepackage{graphicx,epsfig,dcolumn,bm,mathrsfs,amsmath,amssymb}
\usepackage{color}

\graphicspath{{images/}} 

\newcommand{\ud}{\mathrm{d}}
\newcommand{\change}[1]{\textcolor{black}{#1}}

\begin{document}

\title{Elastic Property of Membranes Self-assembled from Diblock and Triblock Copolymers}

\author{Rui Xu}
\affiliation{Department of Physics \& Astronomy, McMaster University, 1280 Main Street West, Hamilton, Ontario, Canada L8S 4M1}

\author{Ashkan Dehghan}
\affiliation{Department of Physics \& Astronomy, McMaster University, 1280 Main Street West, Hamilton, Ontario, Canada L8S 4M1}

\author{An-Chang Shi}
\email[]{shi@mcmaster.ca}
\affiliation{Department of Physics \& Astronomy, McMaster University, 1280 Main Street West, Hamilton, Ontario, Canada L8S 4M1}

\author{Jiajia Zhou}
\email[]{jjzhou@buaa.edu.cn}
\affiliation{Center of Soft Matter Physics and its Applications, Beihang University}
\affiliation{School of Chemistry, Key Laboratory of Bio-Inspired Smart Interfacial Science and Technology of Ministry of Education, Beihang University}
\affiliation{Beijing Advanced Innovation Center for Biomedical Engineering, Beihang University, Beijing 100191, China}


\begin{abstract}
The elastic property of membranes self-assembled from AB diblock and ABA triblock copolymers, as coarse-grained model of lipids and the bolalipids, are studied using the self-consistent field theory (SCFT). 
Specifically, solutions of the SCFT equations, corresponding to membranes in different geometries (planar, cylindrical, spherical, and pore) have been obtained for a model system composed of amphiphilic AB diblock copolymers and ABA triblock copolymers dissolved in A homopolymers. 
The free energy of the membranes with different geometries is then used to extract the bending modulus, Gaussian modulus, and line tension of the membranes. 
The results reveal that the bending modulus of the triblock membrane is greater than that of the diblock membrane. 
Furthermore, the Gaussian modulus and line tension of the triblock membrane indicate that the triblock membranes have higher pore formation energy than that of the diblock membranes.
The equilibrium bridging and looping fractions of the triblock copolymers are also obtained.  Implications of the theoretical results on the elastic properties of biologically equivalent lipid bilayers and the bolalipid membranes are discussed. 
\end{abstract}


\maketitle

\section{Introduction}
\label{sec:introduction}

Amphiphilic molecules are molecules with hydrophilic and hydrophobic components.
When placed in water, these molecules self-assemble into mesoscopic structures such as micelles and vesicles.
In particular, amphiphilic lipids and block copolymers could self-assemble into planar structures in the form of membranes. 
These membranes have a unique combination of properties: they are highly flexible, yet their structural integrity is maintained under strong deformation \cite{Lipowsky1998}. 
These properties are important for many biological processes such as pore and vesicle formation. 
The performance of the self-assembled membranes depends crucially on their elastic property, which is in turn quantified by the elastic parameters, specifically the bending modulus $\kappa_M$, the Gaussian modulus $\kappa_G$, and the line tension of a membrane edge $\sigma$, of the membranes. Therefore it is desirable to understand how the elastic property of self-assembled membranes depends on the molecular details of the self-assembling amphiphilic molecules. 

In this article, we focus on the effect of the architecture of the membrane-forming amphiphilic molecules on the elastic property of the self-assembled membranes.
Specifically, we are interested in comparing the elastic properties of membranes composed of phospholipids to those of the bolalipids. 
The phospholipid has one polar or hydrophilic head group connected to hydrophobic fatty acid chains, which could be represented by an AB diblock copolymer where the A block is hydrophilic to the A solvents and the B block is hydrophobic. 
On the other hand, the bolalipid is a bipolar molecule, consisting of two hydrophilic head groups connected by hydrophobic tail groups \cite{Gliozzi2002, Koga2007}. 
A bolalipid could be represented by an ABA triblock copolymer obtained by connecting two AB diblock copolymer chains at the B-ends.

Bolalipids are exclusively found in archaea, which can survive in harsh environments such as hot springs and salty lakes.
It has been hypothesized that the bolalipids could enhance the rigidity of the membranes, thus allowing the archaeal membrane to maintain its physiological functions in high temperature environments \cite{Oger2013, vandeVossenberg1998, Koga2007, DeRosa1986}. 
However, the origin of this purported increase in rigidity has not been fully understood. 
There are also chemical differences between the phospholipid and the bolalipid: the bolalipid has isoprenoid fatty acid chains that may contain cyclopentane groups. 
Archaea also differ in that they have an S-layer, a 2D protein matrix that may confer additional stability to the archaeal membrane. 
Taking together, it is believe that  these chemical and physical characteristics allow archaea to survive in extreme environments \cite{Oger2013}. 
In the current study, we adopt a simple model composed of AB diblock copolymers or ABA triblock copolymers to represent the topologies of the phospholipids and bolalipids. 
Our model includes the important physical properties of the amphiphilic molecules such as the hydrophilic/hydrophobic interactions in AB diblock copolymer membranes and ABA triblock copolymer membranes, as well as the topological constraints of the lipids.

Many experiments have been performed on phospholipid membranes to determine their bending modulus $\kappa_M$. 
The bending modulus could be obtained from a variety of experimental techniques \cite{Schneider1984, Faucon1989, Rheinstaedter2006, Bo1989, Cuvelier2005}. 
On the other hand, experimental study of bolalipid membranes have been sparse, possibly due to the high cost and scarcity of bolalipids \cite{Bode2008}. 
In the experiment of Duwe et al., it was found that the addition of small percentage of bolalipids (2--5 $\%$ mol) in a phospholipid bilayer would result in a significant reduction in the mean bending modulus \cite{Duwe1990}. 
Furthermore, model bolaamphiphilies have been synthesized to study archaeal bolalipid membranes \cite{Berkowitz1993, Yamauchi1993}, and these authors have found significantly improvement of the membrane performance, specifically in impermeability to ions, in acid tolerance, and in thermal stability.

Theoretically, the elastic properties of lipid membrane have been examined using self-consistent field theory based on the continuous Gaussian chain model \cite{Katsov2004, Katsov2006, Ting2011, LiJianfeng2013b, Dehghan2015} and lattice Scheutjens-Fleer model \cite{Pera2014, Pera2015, Varadharajan2018}, but the studies on the bolalipid membranes have been sparse.
For example, Mukhin and Kheyfets used an analytic approach to compare the free energy per bolalipid with the free energy per phospholipid pair, and found that the free energy per bolalipid was higher than the free energy per phospholipid pair \cite{Mukhin2010}. 
However, these results disagreed with a mean-field molecular theory developed by Longo et al. \cite{Longo2007}. 
Furthermore, molecular dynamics simulations have been used to show how the elastic parameters of bolalipid membrane depend on the presence of methyl groups and cyclopentane rings on the hydrophobic tail groups of the bolalipid \cite{Bulacu2012, Chugunov2014}. 
Despite the very limited amount of studies available in the literature, some consensus on the effect of bolalipids has emerged in that most researchers believe that the bolalipid confers additional stability to the archaeal membrane. 
However, different lipid models have led to some contradicting results. 
In the current work, we compare two membranes where the \textit{only} difference between them is the addition of a constraint to force pairs of AB diblock copolymers to become ABA triblock copolymers. 
Therefore our study sheds light on the effects of molecular topology on the elastic property of the self-assembled membranes.

In order to directly calculate the elastic parameters of the membranes self-assembled from AB diblock copolymers and ABA triblock copolymers, we extend the method developed in our earlier work for AB diblocks to the ABA triblock system \cite{LiJianfeng2013b, Dehghan2015}.  
We use the self-consistent field theory (SCFT) formulated in different geometries to numerically calculate the elastic moduli and the line tension of the membrane. 
Our model system is composed of AB diblock copolymers or ABA triblock copolymers dissolved in A-homopolymers, in which the block copolymers can self-assemble to form membranes. 
We adjust the chemical potentials of the copolymers to form a tensionless planar membrane and then construct tensionless cylindrical and spherical membranes. 
The sizes of the spherical and cylindrical membranes are fixed by a constraint and a constraint is also used to form pores of different radii in a planar membrane. 
The free energy of the membranes in different geometries is then obtained by solving the SCFT equations. 
We then fit the SCFT free energies of these membranes to the Helfrich model to obtain the elastic parameters. 
Within the Helfrich model, the membrane is represented by an elastic sheet \cite{Canham1970, Helfrich1973, Evans1974}, with a free energy given by
\begin{equation}
F = \int \big[2 \kappa_M (M - c_0)^2 + \kappa_G G\big] \ud A + \int \sigma \ud L
\end{equation}
where $M = (c_1+c_2)/2$ is the mean curvature, $G = c_1c_2$ is the Gaussian curvature, and $c_1$ and $c_2$ are the principle curvatures of the membrane. 
In the following, we obtain elastic parameters from SCFT free energies and focus on the differences between the AB diblock and ABA triblock copolymer membranes for different hydrophilic volume fractions $f_A$. 

In addition to the elastic parameters of the ABA triblock copolymer membrane, we calculate the fractions of the copolymers in the looping and bridging configurations. 
The bridging and looping fractions determine the strength of the coupling between the two leaflets of the membrane. A membrane with no looping has leaflets that are completely coupled and can be considered as a monolayer membrane, while a membrane with 100$\%$ looping has leaflets that are decoupled and can be considered to be a true bilayer membrane. 
Processes such as phase separation into domains and shape changes in phospholipid membranes generally involve leaflets that slide relative to one another. 
However, in a membrane where the leaflets are strongly coupled, interleaflet sliding could be suppressed. 
The information about the looping fraction would provide some hints on the dynamics of bolalipid membranes.
Previous works have suggested that the rigidity of the hydrophobic tails can significantly increase the bridging fraction, thus increasing the coupling between leaflets \cite{Bulacu2012}. 
In our model, we represent the copolymers with Gaussian chains and calculate bridging and looping fractions at equilibrium for various curved and flat membranes. 
Recently, Galimsyanov et al. published a set of analytic equations for the elastic deformations of bolalipid membranes, finding asymmetrical distributions of looped bolalipids between leaflets in curved membranes \cite{Galimzyanov2016}. 
Our model allows us to explicitly calculate this asymmetry as a function of membrane curvature.

The remainder of this paper is organized as follows: Section~\ref{sec:model} describes the SCFT model of AB diblock and ABA triblock copolymer membranes and the geometric constraints used in the study. 
Our results on the elastic properties of the membranes and the looping versus bridging configurations of the triblock copolymer are presented in section~\ref{sec:results}.
We conclude in section~\ref{sec:summary} with a brief summary.

\section{Model and Method}
\label{sec:model}

In this section, we first present a brief introduction to the theoretic framework for the free energy calculation of diblock and triblock copolymer membranes.
The general theory of SCFT has been well documented in several excellent reviews and monographs \cite{Schmid1998, Fredrickson2002a, Matsen2002, Shi2004_chapter, GompperSchick1, Fredrickson}, and we refer readers to them for details.
Here we shall focus on the implementation of SCFT in curved geometries, which are essential to obtain the elastic parameters.

\subsection{Free energy of membrane}
\label{sec:freeE}

We consider the membranes composed of AB diblock copolymers and ABA triblock copolymers. 
Both membranes are solvated by A homopolymers. 
The volume fraction of hydrophilic A block in the copolymers is denoted by $f_A$. 
The whole system composed of block copolymers and homopolymers is contained in a volume, $V$.
We assume that the homopolymer and the diblock copolymer have the same degree of polymerization $N$. 
The triblock copolymer has a degree of polymerization $2N$, which is twice of that of the diblock copolymers.
Therefore, the triblock architecture is identical to that of two diblock copolymers connected at the ends of their hydrophobic tails, and the ratio of the triblock to diblock copolymer length is $\kappa = 2$. 
We assume the A and B monomers to have the same monomer density $\rho_0$ and Kuhn length $b$. 
The interaction between A and B monomers is governed by the standard Flory-Huggins parameter, $\chi$. 
The system is further assumed to be incompressible. 
We will present the theory in the grand canonical ensemble, controlled by relative chemical potentials, and we use the homopolymer chemical potential as reference. 
The controlling parameters are the chemical potentials of the diblock, $\mu_d$ and the triblock, $\mu_t$, or the respective activities, $z_d = \exp(\mu_d)$ and $z_t = \exp(\kappa \mu_t)$.

We parameterize the contours of the copolymers using $s$, which increases from 0 to 1 for the diblock copolymers and from 0 to 2 for the triblock copolymers. 
A function $\nu (s)$ is used to specify the nature of the copolymers. 
Specifically, we define the AB diblock copolymer by,
\begin{equation} \label{EQ:diblock}
 \nu (s) =  \begin{cases} 
      A & \text{if} \quad 0<s<f_A \\
      B & \text{if} \quad f_A<s<1
   \end{cases}
\end{equation}
Similarly, we define the symmetric ABA triblock copolymer by,
\begin{equation} \label{EQ:triblock}
 \nu (s) =  \begin{cases} 
      A & \text{if} \quad 0<s<f_A \\
      B & \text{if} \quad f_A<s< 2-f_A \\
      A & \text{if} \quad 2-f_A<s<2
   \end{cases}
\end{equation}

The grand free energy of the system within the mean field approximation has the following form,
\begin{eqnarray}
  \frac{N \mathscr{F}}{k_B T \rho_0}  
  &=& \int \ud \textbf{r} \bigg\lbrace \chi N \phi_A (\textbf{r}) \phi_B(\textbf{r}) 
      - \omega_A (\textbf{r}) \phi_A (\textbf{r}) \nonumber \\
  &-& \omega_B(\textbf{r}) \phi_B(\textbf{r}) 
      - \xi(\textbf{r}) [ 1 -\phi_A (\textbf{r})  - \phi_B(\textbf{r})] \nonumber \\
  &-& \psi \delta (\textbf{r} - \textbf{r}_1) 
      [\phi_A (\textbf{r})  - \phi_B(\textbf{r})] \bigg\rbrace \nonumber \\
  &-& z_d Q_d - \frac{z_t}{\kappa}  Q_t - Q_h
\end{eqnarray}
where $\phi_{\alpha} (\textbf{r})$ and $\omega_{\alpha} (\textbf{r})$ are the local concentrations and mean fields of the $\alpha$-type monomers, where $\alpha = A,B$. 
We introduce a Lagrange multiplier, $\xi (\textbf{r})$, to enforce the incompressibility condition.
We also introduce a second Lagrange multiplier, $\psi$, to pin the membrane at a prescribed location $\mathbf{r}_1$, enforced by the delta function $\delta (\mathbf{r}-\mathbf{r}_1)$.
The quantity $Q_{\beta}$ represents the single chain partition function, with $\beta=d, t$ or $h$ for the diblock, triblock copolymer, and homopolymer, respectively.
These single chain partition functions can be defined in terms of end-integrated propagators $q_{\beta} (\mathbf{r},s)$,
\begin{eqnarray}
Q_d &=& \int \ud \textbf{r} \ q_d (\textbf{r},1) \, , \\
Q_t &=& \int \ud \textbf{r} \ q_t (\textbf{r},2) \, , \\
Q_h &=& \int \ud \textbf{r} \ q_h (\textbf{r},1) \, .
\end{eqnarray} 
These end-integrated propagators satisfy the following modified diffusion equation,
\begin{equation} \label{EQ:Difusion_EQ}
\frac{\partial}{\partial s} q_{\beta} (\textbf{r},s)  = R_g^2 \nabla^2 q_{\beta} (\textbf{r},s) - \omega_{\small{\small{\nu(s)}}} (\textbf{r}) q_{\beta} (\textbf{r},s)
\end{equation}
where $R_g = b \sqrt{N/6}$ is the radius of gyration. 
The mean fields $\omega_{\nu(s)}$ are specified by Eq.~(\ref{EQ:diblock}) and (\ref{EQ:triblock}).
For A homopolymer, the mean field $\omega_{A}$ is used.
The initial conditions for all propagators are $q_{\beta}(\textbf{r},0) = 1$. 
While the triblock copolymer and the homopolymer are symmetrical, the diblock copolymer is not. 
We need a complementary propagator, $q_d^{\dagger} (\textbf{r},s)$ that satisfies the modified diffusion equation with the right-hand side multiplied by -1, and with the initial condition $q_d^{\dagger} (\textbf{r},1) =1$.

The SCFT method involves a mean field approximation where the free energy is calculated using a saddle-point technique such that the functional derivatives of the free energy must be zero,
\begin{equation}
\frac{\delta \mathscr{F} }{\delta \phi_{\alpha}} = \frac{\delta \mathscr{F} }{\delta \omega_{\alpha}} = 
\frac{\delta \mathscr{F} }{\delta \xi} = 
\frac{\delta \mathscr{F} }{\delta \psi} = 0.
\end{equation}
This approximation leads to the mean field equations for concentrations, mean fields, incompressibility, and the pinning constraint. 
The following equations are for the concentrations,
\begin{align}
\phi_A (\textbf{r}) &= \int_0^1 \ud s \ q_h (\textbf{r},s) q_h (\textbf{r},1-s) \nonumber \\
&+ z_d \int_0^{f_A} \ud s \ q_d (\textbf{r},s) q_d^{\dagger} (\textbf{r},s) \nonumber \\
&+ \frac{z_t}{\kappa} \int_0^{f_A} \ud s \ q_t (\textbf{r},s) q_t (\textbf{r},2-s) \nonumber \\
&+ \frac{z_t}{\kappa} \int_{2-f_A}^{2} \ud s \ q_t (\textbf{r},s) q_t (\textbf{r},2-s) \\
\phi_B (\textbf{r}) &= z_d \int_{f_A}^1  \ud s \ q_d (\textbf{r},s) q_d^{\dagger} (\textbf{r},s) \nonumber \\
&+ z_t \int_{f_A}^{2-f_A} \ud s \ q_t (\textbf{r},s) q_t (\textbf{r},2-s)
\end{align}
The mean-fields are in turn determined by the following equations,
\begin{align}
\omega_A (\textbf{r}) &= \chi N \phi_B(\textbf{r}) + \xi(\textbf{r}) - \psi \delta (\textbf{r}-\textbf{r}_1), \\
\omega_B (\textbf{r}) &= \chi N \phi_A(\textbf{r}) + \xi(\textbf{r}) + \psi \delta (\textbf{r}-\textbf{r}_1).
\end{align}
Finally, the incompressibility and pinning constraints are given by,
\begin{align}
&1 = \phi_A (\textbf{r}) + \phi_B(\textbf{r}), \\
&\phi_A (\textbf{r}_1) = \phi_B (\textbf{r}_1). 
\end{align}
This set of equations are solved self-consistently using an iterative algorithm. 

We are interested in the difference in free energy between a membrane system and a reference homogeneous system. 
A homogeneous system has no spatial dependence, and therefore the self-consistent fields become constant (scalar) fields. 
The self-consistent equations can then be solved analytically to determine the single chain partition functions, and average concentration ($\overline{\phi}_A, \overline{\phi}_B$) and mean fields ($\overline{\omega}_A, \overline{\omega}_B$). 
The bulk free energy, $\mathscr{F}_\text{bulk}$ is given by
\begin{equation}
\frac{N \mathscr{F}_{\text{bulk}} }{k_B T \rho_0} = \chi N \overline{\phi}_A \overline{\phi}_B - \overline{\omega \vphantom{+1}}_A \overline{\phi}_A - \overline{\omega \vphantom{+1}}_B \overline{\phi}_B - z_d Q_d -  \frac{z_t}{\kappa}Q_t - Q_h.
\end{equation}
We define the excess free energy density per unit area of the membrane, $F$, by the following equation,
\begin{equation}
F = \dfrac{N (\mathscr{F} - \mathscr{F}_{\text{bulk}})}{k_B T \rho_0 A},
\end{equation}
where $A$ is the area the membrane system.

\subsection{Bridge/loop fraction}

\begin{figure}[htp]
  \centering
  \includegraphics[width=0.8\columnwidth]{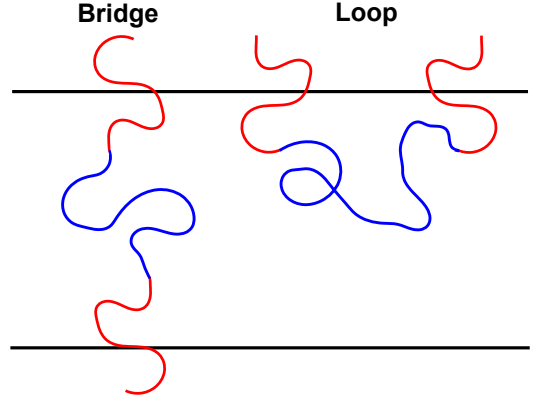}
  \caption{An ABA triblock copolymer in \textit{bridge} configuration on the left, and in \textit{loop} configuration on the right. }
  \label{fig:bridge_loop}
\end{figure}

Different from the AB diblock copolymers, the ABA triblock copolymers could assume two different configurations, i.e. the looping and bridging configurations. 
That is, a triblock copolymer can either span through the membrane in a \textit{bridge} conformation, or it can loop back on itself in a \textit{loop} configuration. 
These two conformations are illustrated in Fig.~\ref{fig:bridge_loop}. 

We wish to determine the fraction of triblock copolymers that exist in the loop versus the bridge conformation \cite{Matsen1999a}. 
In the first step, we define a propagator with their first AB junction confined to the left half of the computation box, $\overline{q}_t^{\,l}$, and a propagator with its first AB junction confined to the right half of the computation box, $\overline{q}_t^{\,r}$,
\begin{align}
\overline{q}_t^{\,l} (\textbf{r}, f_A) &= \begin{cases}
q_t(\textbf{r},f_A) \ &\text{if} \ \ \textbf{r} < \textbf{r}_{mid} \\
0 \  &\text{otherwise}
\end{cases}\\
\overline{q}_t^{\,r} (\textbf{r}, f_A) &= \begin{cases}
q_t(\textbf{r},f_A) \ &\text{if} \ \ \textbf{r} > \textbf{r}_{mid} \\
0 \  &\text{otherwise}
\end{cases}
\end{align}
We then propagate the solution for $s \geq f_A$, and obtain the probability distributions $\overline{\rho}^{\,l} (\textbf{r},s)$ and $\overline{\rho}^{\,r} (\textbf{r},s)$ for the triblock starting from the left half and right half of the computation box, respectively,
\begin{align}
  \overline{\rho}^{\,l} (\textbf{r}, s) &= \frac{1}{Q_t} \overline{q}_t^{\,l} (\textbf{r},s) q_t(\textbf{r},2-s)\\ 
  \overline{\rho}^{\,r} (\textbf{r}, s) &= \frac{1}{Q_t} \overline{q}_t^{\,r} (\textbf{r},s) q_t(\textbf{r},2-s)
\end{align}

The loop probabilities for a polymer that originates from the left half, $\upsilon_L^l$, and the right half, $\upsilon_R^r$, are the integrals of the probability distributions over the respective computational box,
\begin{align}
\upsilon_L^l &= \frac{1}{V_{\rm LHS}} \int_{\rm LHS} \ud \mathbf{r} \, \overline{\rho}^{\,l} (\textbf{r}, 2-f_A)\\
\upsilon_L^r &= \frac{1}{V_{\rm RHS}} \int_{\rm RHS} \ud \mathbf{r} \, \overline{\rho}^{\,r} (\textbf{r}, 2-f_A)
\end{align}
The total loop probability, $\upsilon_L$, and total bridge probability, $\upsilon_B$ are,
\begin{equation}
  \upsilon_L = \frac{1}{2} (\upsilon_L^l + \upsilon_L^r), 
  \quad \quad \upsilon_B = 1- \upsilon_L .
\end{equation}

\subsection{Geometrical constraints}
\label{sec:geometry}

In order to gain information about the membrane's elastic properties, we constraint the membrane to a curved shape and calculate the corresponding excess free energy. 
Specifically, we perform the calculations in the following four geometries: 
(1) an infinite planar membrane, 
(2) a cylindrical membrane with a radius $r$, 
(3) a spherical membrane with a radius $r$, and 
(4) a planar membrane with a circular pore of radius $R$.
Due to symmetry, the first three geometries could be reduced to one-dimensional problems. 
The membrane pore can be stabilized in cylindrical coordinates, and this problem can be reduced to a two-dimensional problem by assuming angular symmetry in the azimuthal coordinate. 
Therefore we need solve the modified diffusion equation (\ref{EQ:Difusion_EQ}) in one dimension and in two dimensions. 
We implement a Crank-Nicolson algorithm to solve the one-dimensional modified diffusion equation, and an Alternating-Direction Implicit algorithm to solve the two-dimensional modified diffusion equation \cite{NR3}. 
Furthermore, we use the planar membrane to find the activities $z_d$ and $z_t$ that correspond to a tensionless membrane. 
We then use the cylindrical membrane to calculate the bending modulus, and the spherical membrane to isolate for the Gaussian bending modulus. 
The pore configuration is used to extract the line tension of the membrane edge. 

The membranes must be stabilized in the specific geometries.
We stabilize the membrane with the constraint field $\psi \delta (\textbf{r} - \textbf{r}_1) [ \phi_A (\textbf{r}) - \phi_B (\textbf{r}) ]$, which forces the concentration of hydrophilic A monomers and hydrophobic B monomers to be equal at the position $\textbf{r}_1$. 
We use this constraint to set the radii of the cylindrical and spherical membranes, and the radius of the membrane pore. 
We set the size of the computational box such that the fields reach bulk values at the edges of the box.
This is to ensure the calculation box is large enough so that the elastic properties are not affected by the boundary conditions.

We calculate the excess free energy of four types of membranes, which are denoted by $F^0$ for the planar membrane, $F^C$ for the cylindrical membrane, $F^S$ for the spherical membrane, and $F^P$ for the pore geometry. 
As mentioned previously, we are interested in a membrane with zero surface tension ($F^0=0$), which we find using a secant method to determine the activity $z_d$ or $z_t$ that corresponds to this state for the diblock and triblock membranes, respectively.  

We now consider the curvatures of cylindrical and spherical membranes, where we define a dimensionless curvature $c = d/r$, where $d$ is a reference thickness of the membrane. 
Similar to Katsov et al., we take the thickness of the membrane to be $d = 4.3 R_g$ \cite{Katsov2004}. 
The principal curvatures of a cylindrical membrane are $c_1 = d/r$ and $c_2=0$, resulting the mean curvature $M = d/2r = c/2$ and the Gaussian curvature $G = 0$. 
For a spherical membrane, $c_1 = c_2 = d/r = c$; the mean curvature is $M = d/r = c$ and the Gaussian curvature is $G = d^2/r^2 = c^2$. 
We can now write the modified Helfrich free energy for cylindrical and spherical membrane with zero surface tension, zero spontaneous curvature, and no edge,
\begin{align}
F^C &= 2 \kappa_M M^2 \quad \quad \quad \quad \ = \frac{1}{2} \kappa_M c^2, \\
F^S &= 2 \kappa_M M^2 + \kappa_G G \quad  = (2 \kappa_M + \kappa_G)c^2.
\end{align}
We define a natural unit for the interfacial free energy, $\gamma_{\text{int}}$ between A and B homopolymers in the limit of large $\chi N$, $\gamma_{\text{int}} = b k_B T\sqrt{ (\chi N) / (6N) }$ \cite{Helfand1971}.
To facilitate the comparison we set the $\chi N=30$ in the definition of $\gamma_{\rm int}$.
This allows us to write the bending moduli as dimensionless quantities,
\begin{equation}
\tilde{\kappa}_M = \frac{\kappa_M}{\gamma_{\text{int}} d^2},  \quad \quad \tilde{\kappa}_G = \frac{\kappa_G}{\gamma_{\text{int}} d^2}, 
\end{equation}

In the pore configuration, we form an open pore with a given radius in a tensionless and planar membrane. 
The excess free energy of the pore is proportional to the length of pore edge. 
For a circular pore, the length of the exposed edge is $L = 2 \pi R$. 
The proportionality constant between free energy and pore radius, $\sigma$, is the line tension,
\begin{equation}
F^P = \sigma \, 2 \pi R.
\end{equation}
We define a unit for line tension $\sigma_0 = k_B T \rho_0 d^2 /N$ to get a dimensionless line tension,
\begin{equation}
\tilde{\sigma} = \frac{\sigma}{\sigma_0}.
\end{equation}

We implement the methods described in this section to calculate the elastic parameters of membranes self-assembled from AB diblock copolymers and ABA triblock copolymers, or AB and ABA membranes, for a variety of parameters such as different $\chi N$, different chain fractions $f_A$, different membrane geometries, as well as for membranes consisting of blends of diblock and triblock copolymers.

\section{Results and Discussion}
\label{sec:results}

In this section, we first present numerical results on the characteristics of tensionless membranes self-assembled from AB diblock and ABA triblock copolymers. 
We then show the SCFT results for the elastic moduli, the bending modulus $\kappa_M$ and the Gaussian modulus $\kappa_G$, and discuss their relation to the microscopic molecular parameter $f_A$.
For ABA membranes, we also investigate the looping fraction and demonstrate the effect of the curvature on the looping fraction. 
Finally we study the  membrane in a pore geometry and compute the line tension.


\subsection{Tensionless membranes}

\begin{figure}[htp]
  \includegraphics[width=1.0\columnwidth]{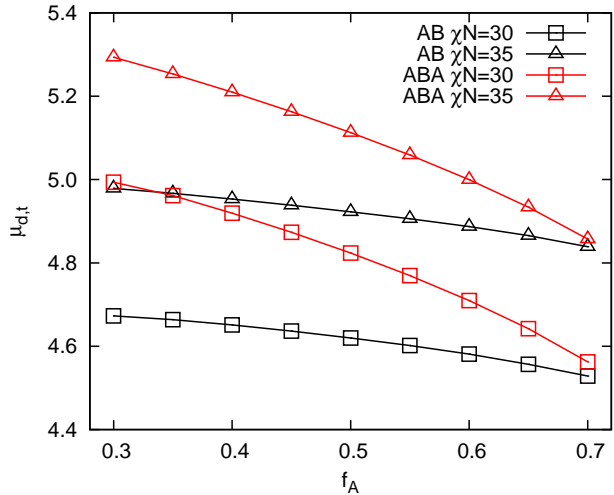}
  \caption{The chemical potential of the diblock and triblock copolymers for a tensionless planar membrane.}
  \label{fig:tensionless}
\end{figure}

In order to obtain tensionless membranes, we adjust the chemical potential $\mu_d$ ($\mu_t$) of the diblock (triblock) copolymers so that the excess free energy of a planar membrane is zero.
Figure \ref{fig:tensionless} shows the chemical potentials corresponding to a tensionless membrane for different $f_A$ values and two interaction parameters $\chi N=30$ and $35$. 
In general, it is found that the chemical potential for tensionless membrane is a decreasing function of $f_A$. 
For large values of $\chi N$, the chemical potential of a tensionless membrane increases.

\begin{figure}[htp]
  \includegraphics[width=1.0\columnwidth]{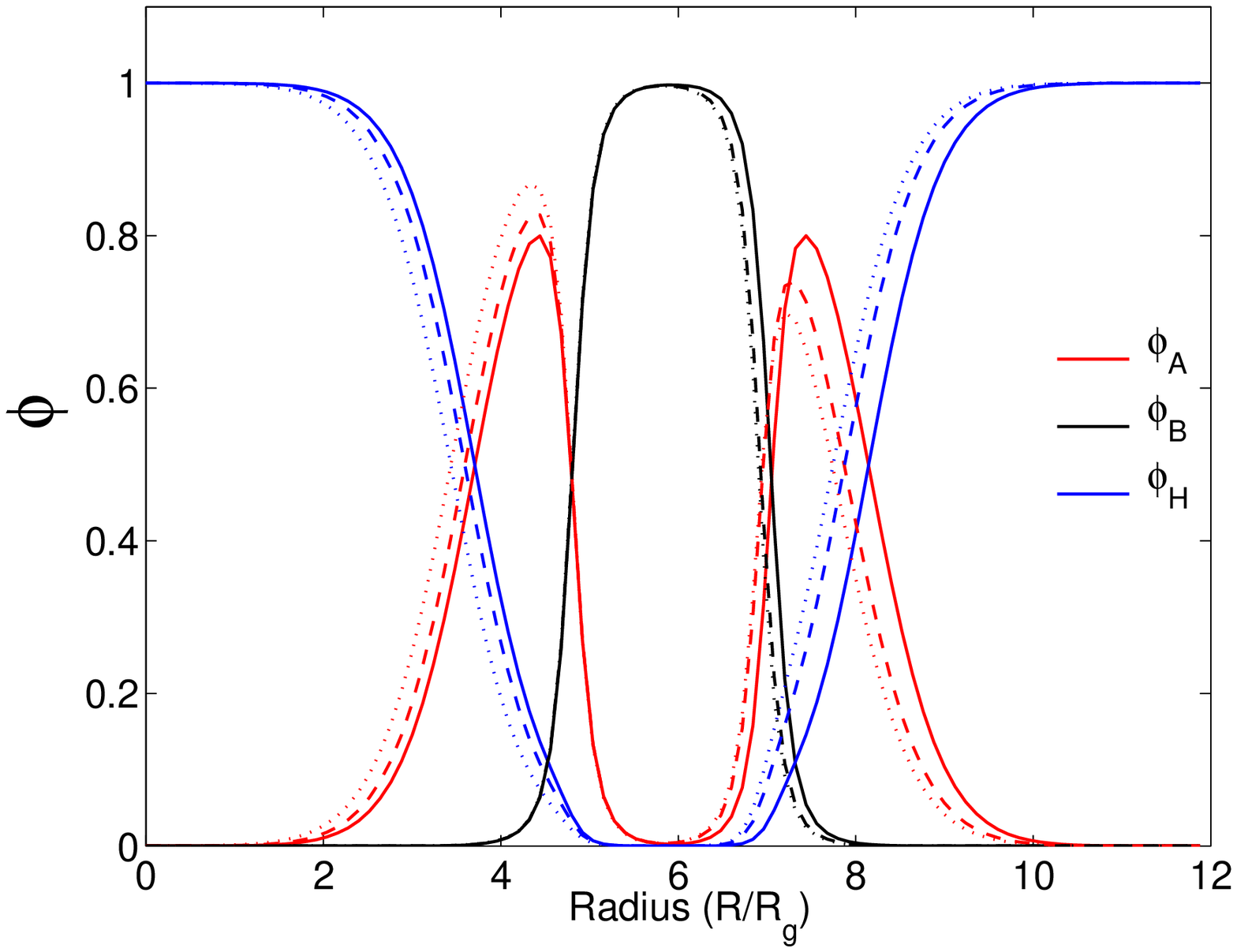}
  \includegraphics[width=1.0\columnwidth]{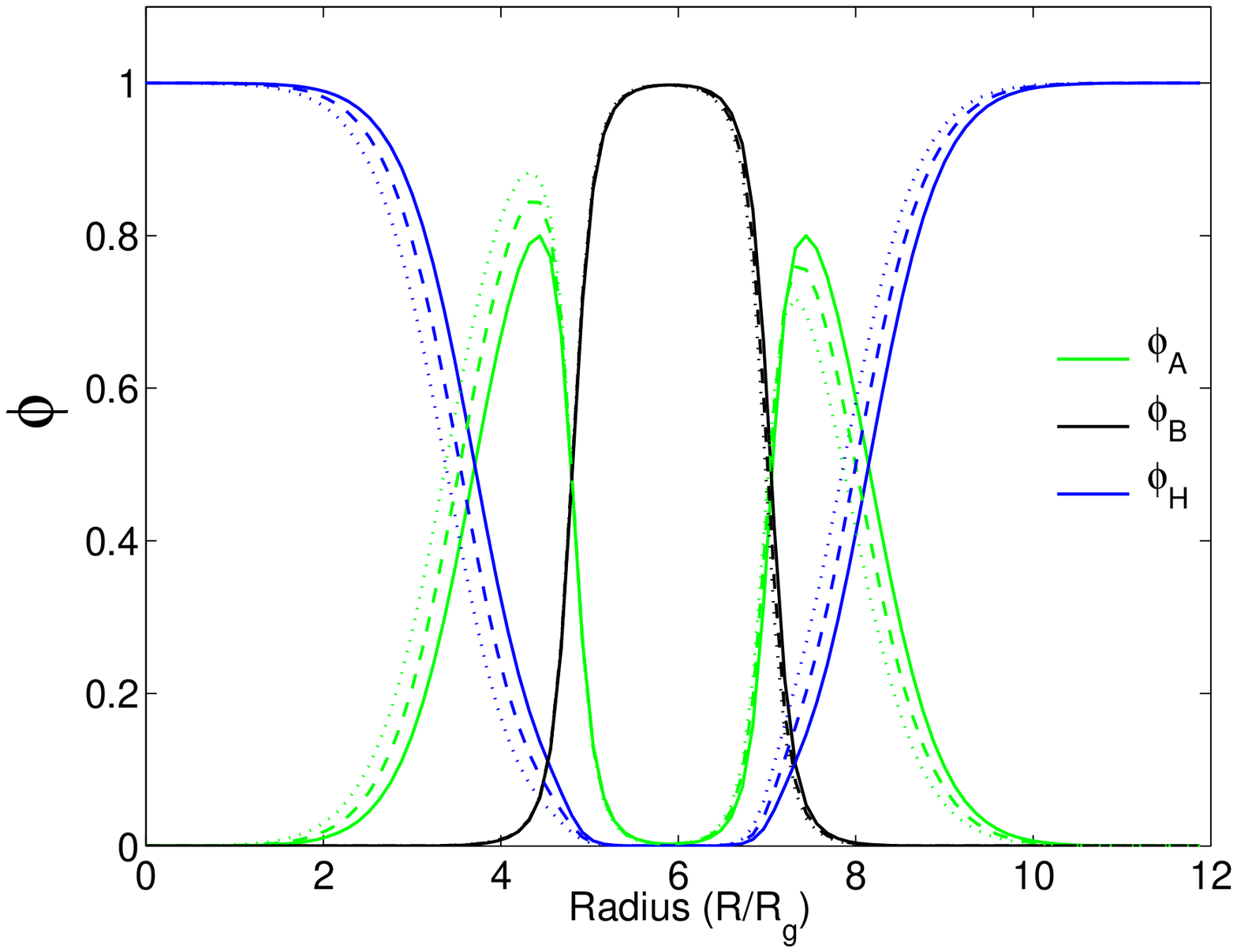}
  \caption{Concentration profiles of (a) AB diblock copolymer membranes and (b) ABA triblock copolymer membranes in planar (solid), cylindrical (dashed), and spherical (dotted) coordinates. The parameters are $f_A=0.5$ and $R = 7R_g$. The asymmetry between leaflets increases from the planar to cylindrical to spherical geometries. The volume fraction of the A-block, B-block, and homopolymer are denoted by $\phi_A$, $\phi_B$, and $\phi_H$, respectively. }
  \label{fig:phi}
\end{figure}

We also show the concentration profiles that correspond to polymeric membranes. 
In Fig. \ref{fig:phi}(a) and (b), we display the density profiles of AB and ABA membranes in planar, cylindrical, and spherical geometries for $f_A = 0.5$, and radius $R = 7R_g$. The position of the membrane within the computational cell is fixed using the auxiliary field $\psi \delta(\mathbf{r}-\mathbf{r}_1)$ to ensure $\phi_A$ = $\phi_B$ at $\mathbf{r}_1$. 
We adjust $\mathbf{r}_1$ to the location that corresponds to the membrane being centered in the computation box.

The concentration profiles show increasing asymmetry between the inner and outer hydrophilic leaflets of the membranes, from the symmetric planar membrane to the highly asymmetric spherical membrane. 
This is due to the smaller surface area of the inner leaflet of the curved membranes, which forces the copolymers to pack more tightly, resulting in a higher local concentration relative to the outer leaflet. 
Another notable asymmetry emerging from these results is a widening of the inner hydrophilic concentration profile and a narrowing of the outer hydrophilic concentration profile. 
This indicates that the copolymers in the inner leaflet are stretched and the copolymers in outer leaflet are compressed relative to the planar membrane. 
These copolymer configurations are entropically unfavorable, resulting in an increase in free energy. 
This increase in free energy is the primary source of the bending energy of the membranes.

\subsection{Bending and Gaussian moduli}

\begin{figure}[htp]
\centering
\includegraphics[width=1.0\columnwidth]{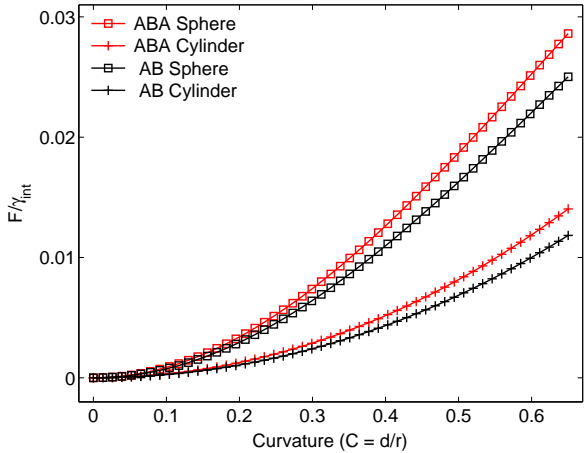}
\caption{Excess free energy for tensionless bilayer (black) and monolayer (red) membranes in cylindrical ($\square$ symbols) and spherical (+ symbols) geometries, as a function of dimensionless curvature $c = d/r$, where $d$ is the thickness of the membrane and $r$ is the radius of curvature. We set the hydrophilic chain fraction $f_A = 0.5$, and immobilize the membrane by pinning the inner leaflet such that the center of the membrane is located at the center of the computation box. }
\label{fig:freeE}
\centering
\end{figure}

In this section, we present our findings for the elastic moduli of triblock copolymer and diblock copolymer bilayer membranes.
We extract the elastic moduli by calculating the excess free energy of tensionless membranes in cylindrical and spherical coordinates, as a function of curvature. 
We then fit the free energy as a function of the curvature to the Helfrich model. 
Figure \ref{fig:freeE} demonstrates the excess free energy for tensionless diblock and triblock membranes with $f_A = 0.5$ in cylindrical and spherical coordinates. 

In Fig. \ref{fig:freeE}, it is immediately apparent that the free energy of the triblock membrane tends to be higher than that of the diblock membrane for all tested curvatures. 
We limit our study to curvatures less than $c = d/r = 0.65$ to avoid contribution from higher order bending terms \cite{LiJianfeng2013b}. 
We then perform a polynomial fit up to second order. 
For a tensionless membrane, the zeroth order term of the polynomial fit should be zero.
This can be seen in Fig. \ref{fig:freeE}, where at zero curvature the excess free energy is also zero. 
For a symmetrical membrane, the linear term in a polynomial fit should also be zero. 
This is demonstrated in Fig. \ref{fig:freeE}, where the slope of the excess free energy curve is zero at zero curvature. 
The quadratic terms, on the other hand, are non-zero and correspond to the bending modulus and the Gaussian modulus. 

\begin{figure}[htp]
\centering
\includegraphics[width=1.0\columnwidth]{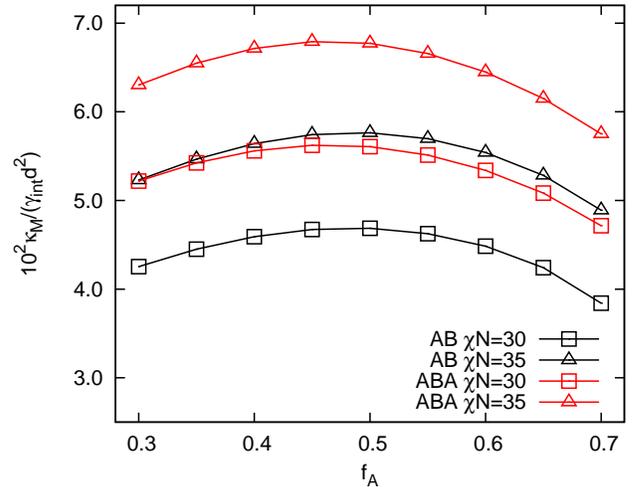}
\caption{Dimensionless bending modulus $\kappa_M$ for bilayer (black) and monolayer (red) membranes as a function of hydrophilic chain fraction, $f_A$. We plot the result for $\chi N = 30$ and $\chi N =35$, with $\square$ and $\triangle$ symbols respectively.}
\label{fig:kappaM}
\centering
\end{figure}

Figure \ref{fig:kappaM} shows that the bending modulus depends weakly on $f_A$, and that the maximum for the diblock membrane occurs near $f_A = 0.5$. 
Interestingly, this maximum is shifted towards lower $f_A$ for the triblock membrane, and occurs closer to $f_A = 0.45$. 
Most interestingly, the bending modulus of the triblock membrane is on average 20$\%$ higher than that of the diblock membrane. 
It is also immediately apparent that an increase in $\chi_N$ results in an increase in $\kappa_M$, without changing its dependence on $f_A$. 
Experimentally, the bending rigidity can be measured using neutron scattering. In Ref. \cite{Choi2002}, Choi et al. studied a non-ionic microemulsion system (C$_i$E$_j$/D$_2$O/n-alkine), and found the bending rigidity increases as the length of C$_i$E$_j$ increases while the hydrophilic-hydrophobic ratio was kept about the same. 
This corresponds to an increase of $\chi N$ in our model and experimental results are consistent with our model prediction. Although in microemulsions the amphiphilic molecules form a monolayer, the dependence of bending modulus on $\chi N$ are similar for the cases of monolayer and bilayer \cite{LiJianfeng2013b}.



\begin{figure}[htp]
\centering
\includegraphics[width=1.0\columnwidth]{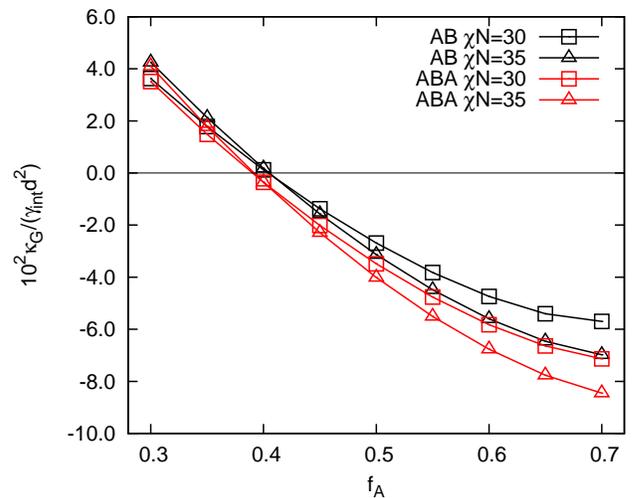}
\caption{Dimensionless Gaussian modulus $\kappa_G$ for bilayer (black) and monolayer (red) membranes as a function of hydrophilic chain fraction, $f_A$. We plot the result for $\chi N = 30$ and $35$, with $\square$ and $\triangle$ symbols respectively.}
\label{fig:kappaG}
\centering
\end{figure}

Figure \ref{fig:kappaG} shows that the Gaussian modulus $\kappa_G$ depends strongly on $f_A$ for both diblock and triblock copolymer membranes.
The value of the Gaussian modulus decreases from positive to negative as $f_A$ increases. 
A negative Gaussian modulus suggests that a spherical geometry is preferred to a planar geometry, while a positive Gaussian modulus suggests the preference of saddle-like geometry with a negative Gaussian curvature. 
We show that $\kappa_G$ for diblock membrane changes sign around $f_A = 0.41$, while $\kappa_G$ for the triblock membrane change sign around $f_A = 0.39$. 
This finding suggests that the triblock membrane is stabilized against pore formation for a greater range of chain architectures than membranes formed from diblock copolymers.
The Gaussian modulus was also measured in the microemulsion systems. 
In Ref.~\cite{Burauer1999}, Burauer et al. measured a negative Gaussian modulus for $f_A \approx 0.5$, and its absolute value increases with increasing $\chi N$. This is consistent with our results.

\begin{figure}[htp]
\centering
\includegraphics[width=1.0\columnwidth]{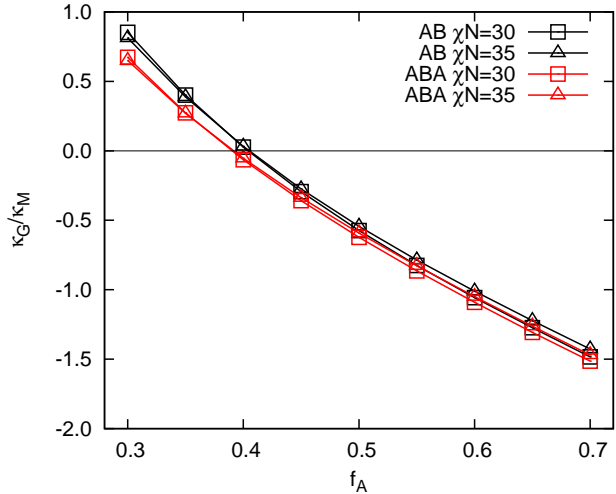}
\caption{Gaussian modulus to bending modulus ratio ($\kappa_G/\kappa_M$) for ABA triblock copolymer (red) and AB diblock copolymer (black) membranes for $\chi N = 30$ and $\chi N = 35$ with $\square$ and $\triangle$ symbols respectively.}
\label{fig:ratio}
\centering
\end{figure}

Figure \ref{fig:ratio} shows that the ratio $\kappa_G$/$\kappa_M$ almost collapses the results for different $\chi N$ onto monotonically decreasing curves from approximately 1 to -1.5. 
This shows that the choice of the interaction parameter has minor qualitative effect on the elastic parameters of the two types of membranes. 
We see that the crossing point, where $\kappa_G/\kappa_M = 0$, remains unchanged from Fig. \ref{fig:kappaG}. 
There is little quantitative difference between $\kappa_G/\kappa_M$ for the triblock membrane and the diblock membrane other than a slight shift in $f_A$ dependence for small $f_A$ values. 

Thus far our results have solely compared the triblock copolymer membrane to the diblock membrane. 
However, we are also interested in membranes consisting of blends of ABA triblock and AB diblock copolymer. 
We define the composition of a blended membrane using the order parameter,
\begin{equation}
  \Omega = \frac{\phi_d -\phi_t}{\phi_d+\phi_t},
\end{equation}
where $\Omega = 1$ corresponds to a pure diblock copolymer membrane and $\Omega = -1$ corresponds to a pure triblock copolymer membrane. 

\begin{figure}[htp]
\centering
\includegraphics[width=1.0\columnwidth]{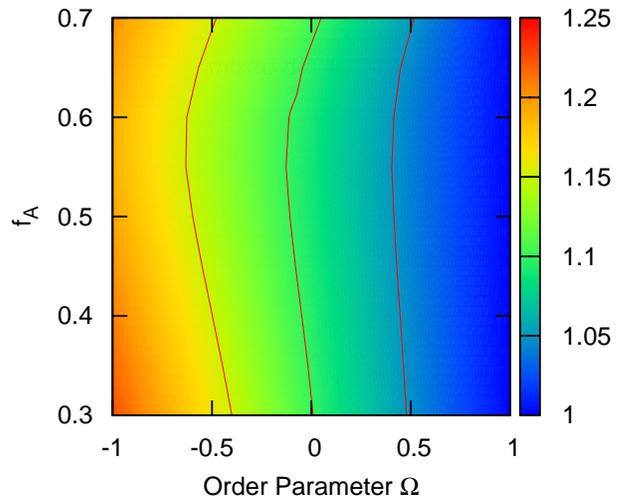}
\caption{Bending modulus of a mixed AB/ABA membrane relative to the bending modulus of an AB diblock copolymer membrane, $\kappa_M/\kappa_M^{d}$. The three contour lines correspond to the value of 1.05, 1.10, and 1.15. We represent the relative proportion of diblock to triblock using the order parameter $\Omega$, where $\Omega = 1$ represents pure diblock copolymer and $\Omega = -1$ represents pure triblock copolymer.}
\label{fig:mix}
\centering
\end{figure}

Figure \ref{fig:mix} shows that the bending moduli of blended ABA triblock and AB diblock copolymer membranes decreases from the ABA bending modulus to the AB diblock modulus approximately as a linear function of the order parameter. 
We also observe that the relative change in bending modulus is slightly higher for lower $f_A$. 
This is another way of representing the $f_A$ shifted $\kappa_M$ maximum of the ABA membrane, shown in Fig. \ref{fig:kappaM}. 
As expected, changing the order parameter from $-1$ to $1$ results in a shift in $\kappa_G$ from the ABA triblock result to the AB diblock result shown in Figure \ref{fig:kappaG}.

\subsection{Looping fraction}

\begin{figure}[htp]
\centering
\includegraphics[width=1.0\columnwidth]{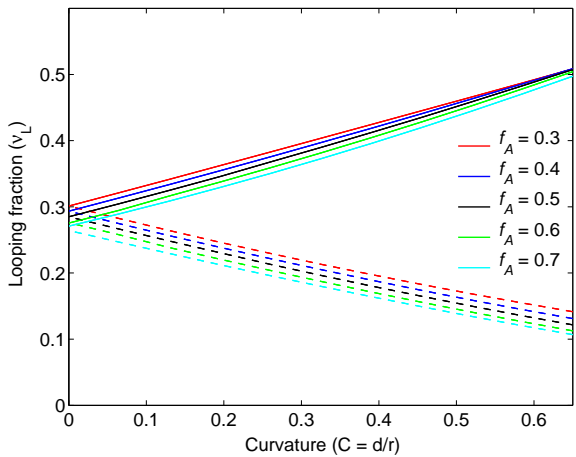}
\caption{Inner (dashed lines) and outer (solid lines) leaflet looping fractions for ABA triblock copolymer membranes as a function of curvature for various $f_A$. We calculate the looping fractions in spherical membrane geometry; the results are qualitatively the same in cylindrical geometry.}
\label{fig:loop}
\centering
\end{figure}

In this section, we investigate the looping fraction in pure ABA triblock copolymer membranes. 
An ABA triblock copolymer can assume two states: the loop state and the bridge state. 
One experimental study on the effect of telechelic triblock copolymers was reported in Ref.~\cite{Maccarrone2014}. The authors found that the bridging can increase the magnitude of both bending and Gaussian moduli. Since only ABA triblock copolymer can form bridges in the membrane, comparison between the moduli of ABA and AB membranes (Fig.~\ref{fig:kappaM} and Fig.~\ref{fig:kappaG}) shows the similar trend.
Looping fractions have also been investigated in the context of lamellar, cylindrical, and spherical ABA triblock bulk morphologies by Matsen and Thompson \cite{Matsen1999a}, who found looping fractions of $55\sim 60\%$, $35\sim 40\%$, and $20\sim 25\%$ for the three respective morphologies. 
We calculate the looping fractions for the inner and outer leaflets of ABA triblock copolymer membranes at different curvatures, and for a range of $f_A$. 
We find looping fractions that are highly dependent on membrane curvature, and the results are shown in Fig. \ref{fig:loop}. 

In Fig. \ref{fig:loop}, we see that the looping fraction of the inner leaflet decreases and the looping fraction of the outer leaflet increases as membrane curvature increases. 
As one would expect, there is no difference in looping fraction between leaflets for planar membranes ($C = 0$). 
We find that the bridge configuration dominates, a result that matches experimental results for bolalipid bridging fractions \cite{Thompson1992, Brownholland2009, Holland2008}. 
Similarly to the concentration asymmetries shown in Fig. \ref{fig:phi}(a) and (b), this finding is due to the relative decrease in inner leaflet area and increase in outer leaflet area as curvature increases. 
The chain architecture $f_A$ also has an effect on the looping fraction, with higher $f_A$ resulting in incrementally lower looping fractions. 

This result shows that the leaflets of the triblock copolymer membrane are strongly coupled, with bridging fractions around 70$\%$. 
This prevents inter-leaflet shear, and may significantly affect the kinetics of processes such as vesicle budding, phase separation, and other morphological membrane changes. 
The dynamics of membranes could potentially be investigated using dynamical version of SCFT \cite{Grzetic2014}.

\subsection{Line tension}

\begin{figure}[htp]
\centering
\includegraphics[width=1.0\columnwidth]{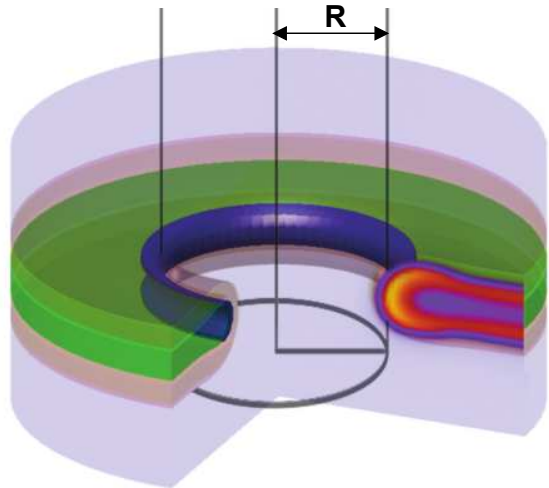}
\caption{We extract line tension by calculating the excess free energy of a pore in a planar membrane, as a function of pore radius. The solvent is blue, the hydrophilic A blocks are red, and the hydrophobic B blocks are green. The inset heatmap shows the A concentration, from  $\phi_A \approx 1.0$ in yellow to $\phi_A \approx 0$ in purple.}
\label{fig:pore}
\centering
\end{figure}

When a pore is formed in a membrane, the lipids reorient themselves to avoid unfavorable interactions between the solvent and the hydrophobic tails. 
An analogous process occurs in diblock and triblock copolymer membranes, where the copolymers reorient themselves to avoid any interface between the homopolymer solvent and the hydrophobic B blocks of the membranes. 
However, this reorientation comes at the cost of a decreasing the configurational entropy of the polymers in the membrane edge. 
This energy penalty is the main contributor to the line tension of the pore. 
We calculate line tension by forming a pore in a planar membrane, and assume that the pore has azimuthal symmetry. 
Figure \ref{fig:pore} demonstrates the pore geometry, and we define the radius $\mathbf{R}$ of the pore as the distance from the center of the pore to the pinned location $r_1$, which is enforced by the Lagrange Multiplier $\psi \delta (\mathbf{r} - \mathbf{r}_1)[\phi_A(\textbf{r}) - \phi_B(\textbf{r})]$. 

\begin{figure}[htp]
\centering
\includegraphics[width=1.0\columnwidth]{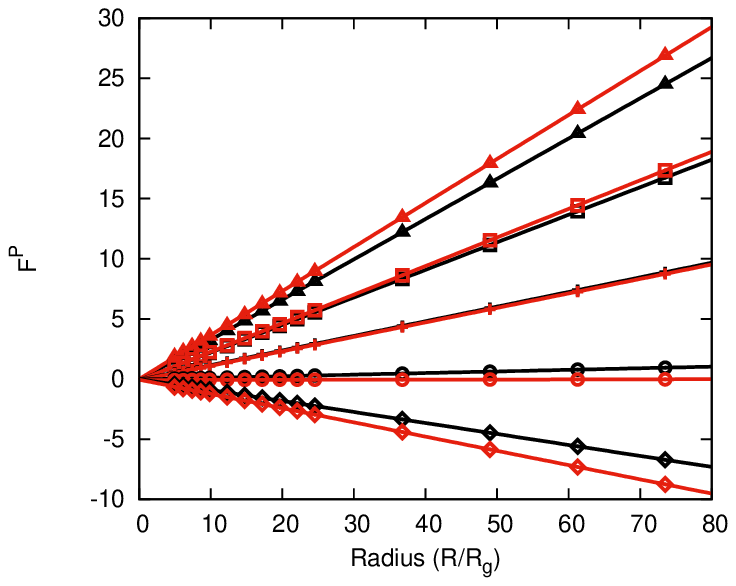}
\includegraphics[width=1.0\columnwidth]{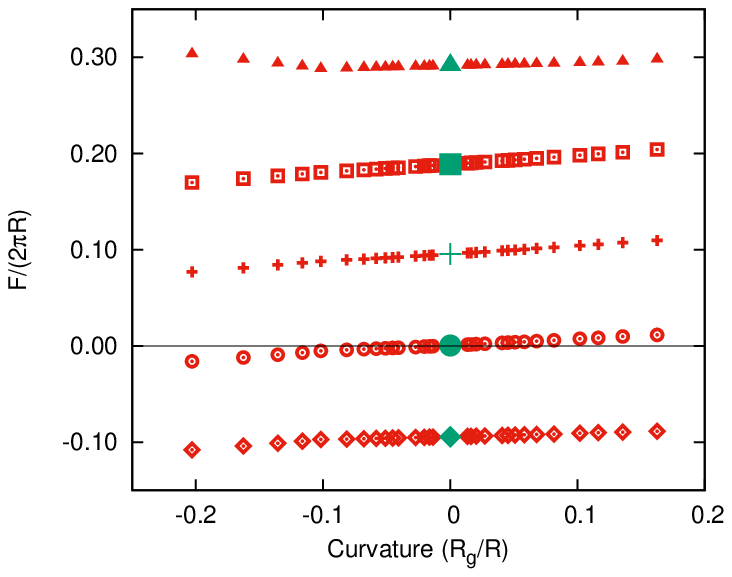}
\caption{\change{(a) Excess free energy for a pore $F^P$ in a planar bilayer membrane as a function of pore radius $R$, for $f_A = 0.3,0.4,0.5,0.6,0.7$, denoted by $\bigtriangleup$, $\square$, $+$, $\circ$, and $\diamond$ respectively. The red symbols are results for triblock membranes and the black symbols for diblock membranes. (b) Excess free energy per unit length of the pore perimeter $F^P/(2\pi R)$ as a function of the curvature of the pore $-1/R$. Also shown are results for a membrane edge (green, curvature zero) and membrane disks (curvature positive).} }
\label{fig:pore_freeE}
\centering
\end{figure}

We begin by calculating the chemical potential corresponding to tensionless planar membranes. 
We then proceed to calculate the excess free energy of pores of different radii. 
\change{Figure \ref{fig:pore_freeE}(a) plots the free energy of a pore $F^P$ as a function of the radius of the pore $R$, for different $f_A$ and for $\chi N = 30$.} 
The line tension for both ABA triblock and AB diblock membranes decrease as $f_A$ increases from positive values at lower $f_A$ to negative values for $f_A$ greater than approximately 0.6.

\change{There are different methods to compute the line tension. 
One way is to fit the data shown in Fig.~\ref{fig:pore_freeE}(a) with a linear least squares fit and the slope gives the line tension $\sigma$.
Another method is to plot the free energy per unit length of the pore perimeter $F^P/(2\pi R)$ as a function of the curvature of the pore $-1/R$ and then extrapolate to curvature zero. 
Fig.~(\ref{fig:pore_freeE}(b) shows the results. 
For completeness, we also carry out two more calculations: one is in the 2-dimension Cartesian coordinate for a membrane edge, which corresponds to the limit of infinitely large radius of the pore, i.e. zero curvature. 
The other one is again in the cylindrical coordinate but for a membrane disk with radius $R$, which corresponds a positive curvature $1/R$. 
For $f_A=0.3$, the curve $F/(2\pi R)$ has a minimum at around $1/R=-0.1/R_g$, indicating there is a preferred curvature of the pore. 
This is consistent with the result in Ref.~\cite{Pera2015}. 
For larger $f_A$ value, the curves are monotonic function of the curvature. }

\begin{figure}[htp]
\centering
\includegraphics[width=1.0\columnwidth]{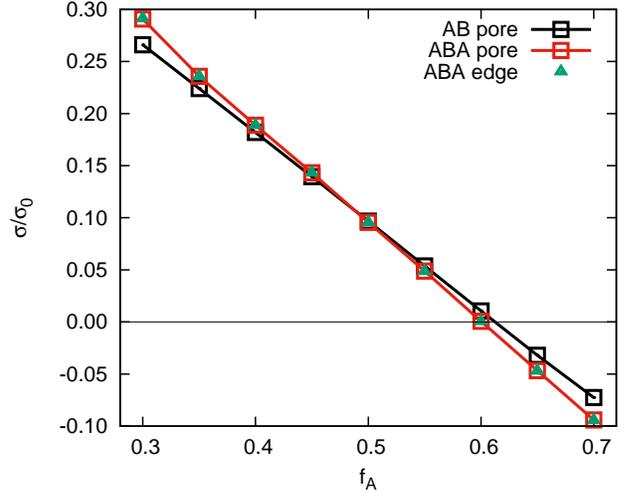}
\caption{Dimensionless line tension for ABA triblock copolymer membranes in a pore geometry (red) and in an edge geometry (green), and AB diblock copolymer membranes (black), calculated from the free energies of pores in planar membranes.}
\label{fig:line}
\centering
\end{figure}

\change{Here we use the first method to be consistent with our previous works.
In Fig.~\ref{fig:line}, we plot the line tension of a triblock membrane as a function of $f_A$.   
The results from the membrane pore and membrane edge are the same. 
For comparison, we also show the line tension of a diblock membrane pore.}

For smaller values of $f_A$, the line tensions for the triblock membrane are larger than that of  diblock membrane.
However, as $f_A$ increases, the difference between the line tensions of two membranes converge, and line tensions are almost the same for $f_A=0.5$.
For large values of $f_A$, the trend is reversed; the line tensions for the diblock membrane are larger than that of triblock membrane.
Also in the range of large $f_A$, line tensions cross the zero point and change from positive to negative value.
For diblock membrane, the condition of $\sigma = 0$ occurs at approximately $f_A = 0.62$, while in the triblock membrane this occurs at approximately $f_A = 0.60$. 
The qualitative behavior of line tension in the two types of membranes were the same. 
In both cases, we have regions of negative line tension corresponding to lipids that stabilize the pores, a result that has been observed experimentally and in molecular dynamics simulations \cite{Karatekin2003, deJoannis2006}. 
However, most lipids have hydrophilic chain fractions less than 0.5, which means that most biological membranes are stabilized against pore formation. 

\section{Summary}
\label{sec:summary}

We have calculated and contrasted the elastic parameters of AB diblock copolymer and ABA triblock copolymer membranes, where the triblock copolymers are obtained by connecting two diblock copolymers at the ends of their respective B blocks. 
We investigated the bending modulus, the Gaussian modulus, and the pore line tension of the two types of membranes using equilibrium self-consistent field theory. 
We found differences in the elastic parameters of the two types of membranes that are solely the result of this connection between B blocks.

A 20$\%$ increase in bending modulus for triblock membranes relative to diblock membranes is predicted by our theory. This is a result largely independent of the volume fraction of the hydrophilic A block, $f_A$. 
By tuning the relative concentration of diblock to triblock copolymer, we showed that bending modulus increases proportionally to the concentration of triblock copolymer. 

We calculated the Gaussian modulus for both membranes and found both to be monotonically decreasing functions of $f_A$, from positive to negative. However, it was found that the triblock membrane had a Gaussian modulus that crossed from positive to negative at $f_A \approx 0.39$, whereas the diblock membrane had a Gaussian modulus that crossed at $f_A \approx 0.41$. This suggests that the triblock membrane is stabilized against pore formation for a larger range of $f_A$ than the diblock membrane.

We calculated the looping fractions in spherical ABA triblock copolymers for different curvatures. At small curvatures, we found looping fractions to be approximately 30$\%$, which is similar to experimental results for bolalipid membranes \cite{Thompson1992, Brownholland2009, Holland2008}. 
As we increased curvature, we found that the looping fraction would decrease in the inner membrane leaflet, and increase in the outer membrane leaflet. 
This suggests that in a dynamically bending membrane, the lateral diffusion and flip-flop rates of the triblock copolymer becomes especially important because unlike the phospholipid membrane, the leaflets of the bolalipid membrane are strongly coupled and cannot slide against each other. 
Although comparisons of interleaflet viscosity are beyond the scope of this paper, this phenomenon could potentially be investigated using dynamical self-consistent field theory \cite{Grzetic2014}.

We calculated the line tension for both membranes, and found similar, monotonically decreasing line tensions as $f_A$ increased. 
We find for small $f_A$ value, which is the biological relevant, the line tension for the triblock membrane is larger than that of diblock membrane.
This result, in combination with the difference in Gaussian modulus,  shows that pore formation is less favorable in a triblock copolymer membrane in terms both the topological transition and the energy penalty to maintain the pore geometry.

The results comparing the elastic parameters of AB and ABA membranes show that the triblock membranes are more rigid and less able to form pores.
We proposed that we could extend these results to biological membranes, in order to compare the properties of phospholipid membranes to bolalipid membranes. 
Biological membranes tend to be extremely complex systems, with an assortment of lipids, proteins, steroids and oligosaccharides. 
However, the most fundamental property of the membrane is the amphiphilicity of its lipid constituents \cite{Singer1972}. 
By representing the phospholipid and the bolalipid as AB diblock copolymers and ABA triblock copolymers and comparing the elastic parameters of the membranes, we can understand the effect of linking to diblock copolymers to form a triblock copolymer membrane will have. 
This is analogous to joining two phospholipids together at the fatty acid tails and then investigating the elastic parameters of the membrane. 
While there are other properties of the bolalipid to consider, such as the isoprenoid chains and cyclopentane rings in bolalipids, these properties have been shown to increase the rigidity of the lipid, which would in turn increase the magnitude of the elastic parameters calculated here \cite{Bulacu2012, Chugunov2014}. 
Therefore, our estimates form a lower limit for any expected change in elastic properties of bolalipid membranes.

It has long been hypothesized that the bolalipid confers additional stability to biological membranes, and that bolalipid membranes play an important role in allowing archaea to survive in hostile environments such as high temperature and high pressure. 
We believe that this extension from amphiphilic block copolymer membrane to phospholipid and bolalipid membranes gives insight into the behavior of both types of membranes, and supports hypothesis that bolalipids are included to increase thermal and mechanical stability in archaeal membranes \cite{Oger2013, vandeVossenberg1998, Koga2007, DeRosa1986}.
Understanding the properties of the bolalipid membrane is also important for various biotechnological applications \cite{Jacquemet2009, Bulacu2012, deVos2014}.

This work has extended the methodology introduced by Li et al. to ABA triblock copolymer membranes, and demonstrated the precision with which the elastic parameters of different amphiphilic block copolymer membranes can be calculated and compared. 
By assuming symmetries in planar, cylindrical, spherical, and pore membrane geometries, we reduce the computational cost of solving the self-consistent equations for these systems to calculations that can easily be performed on a single CPU and that will converge to the equilibrium solution within minutes. 
This framework allows for the membrane parameter space to be investigated quickly and computationally inexpensively. 
\change{One drawback of our model is the lack of molecular details, therefore one must be cautious in the direct comparison between the model prediction and biological value.  
Extensions to charged systems, and incorporation of more biological features from membrane lipids \cite{Pera2014, Pera2015} are feasible, and provide interesting directions for future research.} 
There are many questions that can be answered using this method, and potential future directions include investigating asymmetrical triblock copolymers, and blends of diblock and triblock copolymers with $\kappa \neq 2$. 
Another interesting question is how other constituents of biological membranes such as cholesterol might affect the elastic properties of membranes. 
There are many avenues for using SCFT to calculate the elastic parameters of membranes, making this an exciting and invigorating subject for future work.

 \begin{acknowledgments}
We acknowledge support from the Natural Sciences and Engineering Research Council of Canada (NSERC) and National Natural Science Foundation of China (NSFC) through Grants No. 21504004 and 21774004.
This work was made possible by the facilities of the Shared Hierarchical Academic Research Computing Network (SHARCNET:www.sharcnet.ca) and Compute/Calcul Canada.
 \end{acknowledgments}



\bibliography{membrane}


\end{document}